# ME-FIRST: A Metasurface-Enhanced Fingerprint InfraRed Spectroscopic Tool for Fluid Analytes

XIANGYU ZHAO,[1] YUQING LIU,[1] JINGZHU SHAO, [1] LONGSHENG FANG, [2] AND CHONGZHAO WU [1,*]

[1] *Center for Biophotonics, Institute of Medical Robotics, School of Biomedical Engineering, Shanghai Jiao Tong University, Shanghai, 200240, China*
[2] *School of Materials Science and Engineering, Shanghai Jiao Tong University, Shanghai, 200240, China*
*\*Corresponding Author: czwu@sjtu.edu.cn*

**Abstract:** Infrared (IR) spectroscopy has emerged as a pivotal tool in biomedical diagnostics, offering label-free spectral biomarkers for the detection of numerous diseases, particularly in the fingerprint region. However, the lack of rapid and sensitive IR spectroscopic techniques for analyzing complex fluid analytes remains a critical challenge in clinical practice. To address this limitation, we present a Metasurface-Enhanced Fingerprint InfraRed Spectroscopic Tool (ME-FIRST) that enhances light-matter interactions in sub-wavelength volumes through plasmonic resonances across the entire fingerprint range from 1900 $cm^{-1}$ to 1000 $cm^{-1}$. Numerical simulations reveal confined and enhanced electric near-field, with an average probing depth of ~100 nm and enhancement factor $|E/E_0|$ of ~60-fold at resonant peaks. The ME-FIRST device is further experimentally fabricated and validated, and as a proof of concept, we demonstrate the sensing of molecular vibrational modes with a considerable sensitivity in *L*-lysine over the full fingerprint IR spectral range. The proposed ME-FIRST presents a promising platform for high-sensitivity IR spectroscopy of fluid analytes, paving the way for clinical applications of infrared spectroscopy in biofluid analysis and pathological scenarios.

## 1. Introduction

Infrared (IR) spectroscopy has emerged as a pivotal analytical technique in modern research, owing to its ability to probe molecular vibrations through characteristic absorption of chemical bonds, such as S-H, C-H, N-H, and O-H, with high specificity and non-destructive sampling. [1,2] Enabled by significant advances in Fourier transform infrared (FTIR) instrumentation and analytical methodologies, this technology has found widespread and diverse applications across chemical analysis, environmental monitoring, and the food industry, etc.. [2–6] Particularly in biomedical and clinical research, the fingerprint region (1900-1000 $cm^{-1}$) [1,2] has proven to be invaluable for the efficient detection of natural biomarkers related to disease propagation, including proteins, lipids, nucleic acids, and carbohydrates. [1] Notably, when the high-content fingerprint IR spectra are processed using machine learning (ML) or deep learning (DL) methods, [7] they can bring remarkable advances in clinical pathology and diagnostics. Given this context, infrared spectroscopy has greatly enhanced the understanding of diseases affecting various organs, [8–13], especially those targeting the vibrations of biomarkers in the form of small molecules. [14–16]

Despite these advantages, an effective clinical translation of infrared spectroscopy faces a fundamental limitation: strong absorbance of infrared light by water in fluid specimens. In clinical pathology testing, a substantial proportion of analytes exists in fluid form, such as fine-needle aspiration (FNA) extracts, blood-derived components (e.g., plasma, serum, and whole blood), and various secretions (e.g., saliva, sputum, and bodily exudates). [17] While conventional approaches for fluid infrared spectroscopy require prior sample dehydration to mitigate water interference, [18] such pretreatment inevitably alters native molecular compositions and compromises the integrity and accuracy of resulting spectroscopic data and further diagnosis. [17] Hence, attenuated total reflection-infrared (ATR-IR) has also been implemented for spectroscopy on biofluidic samples. [19] Although ATR-IR achieves a fixed and limited path length via the evanescent field used to probe the substances in the water and can be used to investigate fluid samples, its sensitivity remains inadequate without any field enhancement for stronger light-matter interaction. [20]

Recent advances in resonant electromagnetic nanostructures comprising carefully designed nanostructures have enabled precise control of light confinement, generating strongly enhanced near-fields that significantly amplify light–matter interactions. [21] In the infrared regime, such nanostructures have opened new possibilities through the surface-enhanced infrared absorption (SEIRA) effect. By engineering nanostructures to support localized surface plasmon resonances (LSPRs) or surface plasmon polaritons (SPPs) in engineered nanostructures, SEIRA can be excited through concentrating incident IR light into subwavelength hot spots with ultra-confined and enhanced electric near-field. [22] SEIRA can dramatically amplify the vibrational signals of molecular bonds adsorbed on the surface, enabling ultrasensitive detection, [23] and the confinement of near-field on a nanometer scale can reduce the absorption of water on IR light. SEIRA has shown promise for detecting or monitoring various biomedical fluid analytes, including protein, [24] lipid, [25] polymers, [26] or biological cells. [19] However, traditional SEIRA platforms are often designed with patterns that operate at discrete, narrowband frequencies to match vibrational modes of a small subset of analytes, [27] severely limiting their applications in the real-world complex biological systems requiring broadband analysis, [28] or multi-wavelength IR analysis.

Similar to SEIRA, metasurface-enhanced infrared spectroscopy offers the possibility of offering a broader frequency range of detection by employing a more flexible designed metasurface. [29,30] Several strategies have been proposed to enable broadband, multi-wavelength plasmonic resonances by integrating multiple resonant structures within a single unit cell, each targeting a strong resonance at distinct wavelengths. [25,31] However, when these resonators are closely spaced in resonant frequencies, near-field coupling induces mode hybridization, shifting or weakening the intended resonances, thereby compromising spectral fidelity compared to the desired distribution. This fundamental constraint makes it exceptionally challenging to achieve spectrally tight-spaced resonances required for complete coverage of the narrow infrared fingerprint region. In such a context, polarization-multiplexing offers a solution to this challenge. [28,30] Considering the anisotropy of asymmetric plasmonic nanostructures, their resonant modes are inherently polarization-dependent, demanding alignment between the incident polarization of the electric field and the structures. This polarization-selective behavior enables expanded functionality, such as reconfigurable optical switching. [32] This allows spectrally distinct modes to coexist within the same spatial region when excited by orthogonal polarizations, effectively decoupling adjacent resonances while enabling systematic tuning across the fingerprint spectrum.

In this work, we present a metasurface-enhanced fingerprint infrared spectroscopic tool (ME-FIRST) as a novel and reliable platform for label-free fluid analyte analysis with fingerprint infrared spectroscopy, as shown in **Figure 1a**. Our ME-FIRST features an innovative pattern of the metasurface, achieved by arranging polarization-anisotropic nanostructures with polarization-dependent plasmonic resonances that enhance the light-matter interactions in sub-wavelength volumes across the fingerprint infrared region, as depicted in Figure 1b. First, we demonstrate the polarization-multiplexed coupling between these plasmonic resonances using electromagnetic simulations, and the geometric tunability of the modes is simulated to choose the values of the geometric parameters. We reveal that ME-FIRST can generate strong near-field enhancement (with an average factor $|E/E_0|$ of ~60-fold) confined in the hot spots (with a probing depth of ~100 nm), supporting spatially overlapping yet spectrally distinct plasmonic resonances covering the entire fingerprint infrared region (1900-1000 cm$^{-1}$), as shown in Figure 1c. Second, fabricated via electron-beam lithography (EBL) and electroplating, the ME-FIRST platform demonstrates experimental reflectance characteristics matching theoretical predications. Third, a proof-of-concept testing of molecular vibrational modes in *L*-lysine solutions across different concentrations is implemented. The comprehensive fingerprint spectra of *L*-lysine are successfully achieved, with a considerable sensitivity of 0.131 mg/ml and a limit of detection (LOD) of 0.076 mg/ml. Our method presents an innovative tool supporting plasmonic resonances covering the entire infrared fingerprint region, addressing a critical gap in infrared biosensing. Given the clinical prevalence and significance of liquid biopsies, our technology opens new avenues for infrared spectroscopic analysis in medical diagnostics.

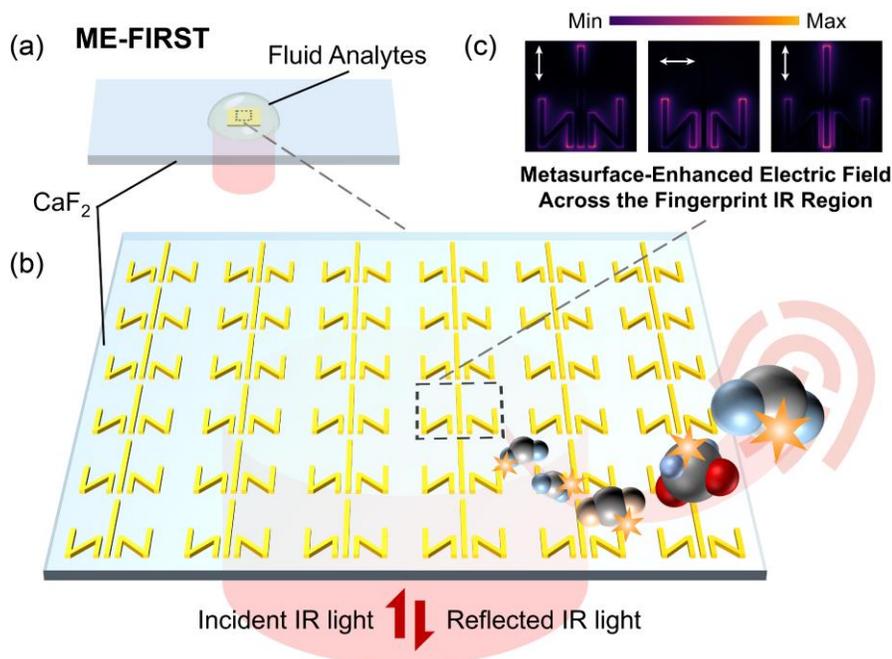

**Figure 1. The sketch map of the metasurface-enhanced fingerprint infrared spectroscopy tool (ME-FIRST).** (a) The calcium fluoride ($CaF_2$) -based ME-FIRST in this work designed for fingerprint infrared detection of fluid analytes. (b) The simplified diagram of the ME-FIRST, consisting of an array of our designed gold nanostructures. The fingerprint IR absorbance spectra are acquired in reflectance mode from the substrate side, capturing light–matter interactions on the other side with nanostructures. (c) The polarization-multiplex electric near-field enhanced by ME-FIRST, enhancing light-matter interactions with multiple resonances across the fingerprint IR region.

## 2. Results and Discussion

*2.1 Design of ME-FIRST and Simulations of Reflectance*

The development of the metasurface-enhanced fingerprint infrared spectroscopic tool (ME-FIRST) requires an engineered metasurface capable of enhancing light-matter interactions across the complete fingerprint spectral range (1900-1000 cm$^{-1}$) through spatially co-localized yet spectrally distinct strong resonance modes. We designed a polarization-multiplexed broadband plasmonic metasurface, with metaunits shown in **Figure 2a**, that simultaneously supports multiple spectrally separated yet spatially overlapping plasmonic resonances by assigning resonant peak responses to different polarization states of incident light. Specifically, the proposed metaunit integrates gold structures of the central nanorod and the side double 'N' proposed by our second author in her previous work. [33]

The full-wave finite-difference time-domain (FDTD) simulations were conducted to characterize the ME-FIRST and analyze its electromagnetic characteristics. Figure 2b presents the simulated reflection spectra of the designed tool under the incident light of both 0° and 90° polarizations. Herein, the ME-FIRST can support reflections of the light in the wavenumber range from 2000 to 1000 cm$^{-1}$, with three peaks $P_0$, $P_1$, and $P_2$ uniformly distributed across the target spectral range.

These resonant modes were systematically tuned by optimizing the geometric parameters shown in Figure 2a. Through comprehensive parameter sweeps, we established the relationship between structural dimensions ($L_1$, $L_2$, $L_3$, and $L_4$) and the spectral response, with values of the structure parameters finally chosen to get the reflectance spectrum matching the requirement of ME-FIRST. The results are shown in Figure 2c-f, and the correlation between the wavelengths of the reflectance peaks and the parameters is demonstrated. The length $L_1$ is found to primarily govern the resonance of the gold nanorod, with increasing values inducing red shifts in $P_0$ and $P_1$ while leaving $P_2$ unaffected. Meanwhile, $P_1$ will not appear unless $L_1$ is more than ~double the length of $L_2$ for the center nanorod to dominate the resonance coupling. The resonance characteristics of double 'N' are collectively controlled by parameters $L_2$, $L_3$, and $L_4$. When $L_2$ increases, the $P_0$ and $P_1$ will move toward longer wavelengths, with their reflectance intensities of resonance decreasing and full-width-at-half-maximums (FWHMs) broadening. Increasing $L_2$ causes longer side nanorods in the double 'N', and two additional strong peaks will appear in the wavenumber range over 1500 cm$^{-1}$ as the side nanorods in the double 'N' begin to dominate the resonance coupling. $L_3$ elongation similarly produces red shifts of $P_0$ and $P_1$. Notably, compared to the double 'N' structure proposed in the previous work, [33] $L_4$ emerged as the critical parameter for resonance peaks positioning. Based on the coupling mechanism of the two structures, extending the outer rod of the double 'N' configuration enlarges its capacitive coupling area with the center nanorod, reinforcing their resonant interaction. As $L_4$ increases from zero, the resonance peak $P_0$ around 1800 cm$^{-1}$ will be developed and intensified. Likewise, elongating $L_2$, $L_3$, or $L_4$ results in a red shift of the $P_2$ resonance peak. The reflectance of the final parameter values, indicated by white dashed lines in the optimization plots, was tuned to achieve optimal spectral coverage, demonstrating how precise geometric control of the coupled nanostructures enables uniform distribution of three distinct resonance peaks across the fingerprint region.

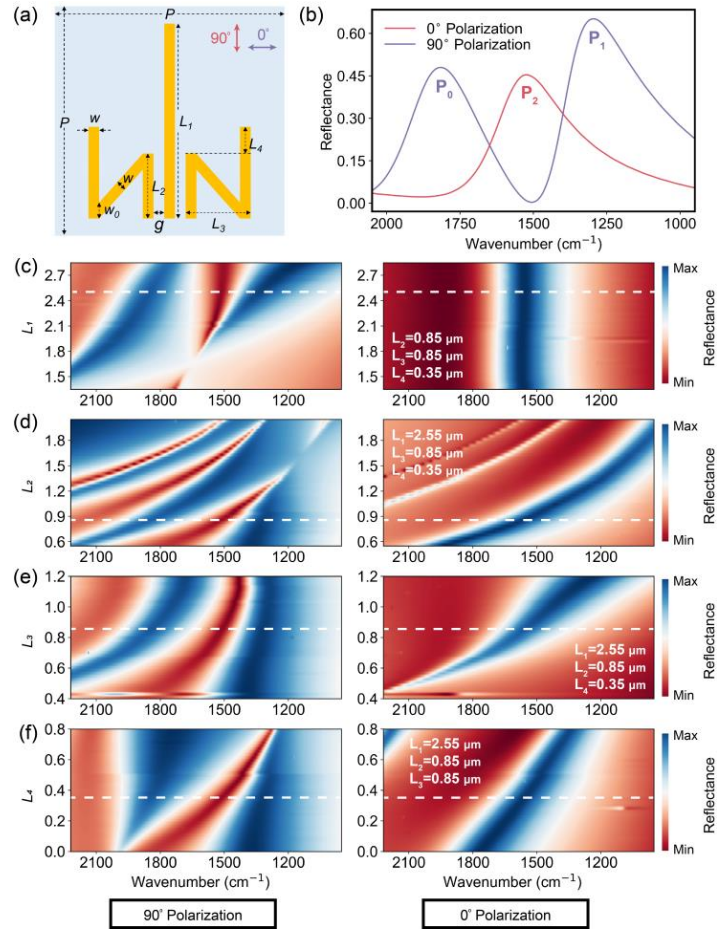

**Figure 2.** The design and simulation results of the metaunit. (a) The geometrical design parameters for the structure of the metaunit. The key parameters are $P = 3$ μm, $L_1 = 2.55$ μm, $L_2 = 0.85$ μm, $L_3 = 0.85$ μm, $L_4 = 0.35$ μm, $w = 0.14$ μm, $g = 0.14$ μm, with $w_0$ changed with $L_2$ and $L_3$ to control the width of the diagonal rods. The thickness of the gold layer is 110 nm. (b) Simulated reflectance of incident infrared light in 0° and 90° polarizations, with the 3 peaks ($P_0$, $P_1$, $P_2$) covering the whole fingerprint region (1900-1000 cm$^{-1}$). (c-f) The parameter sweeping results of $L_1$, $L_2$, $L_3$, and $L_4$, respectively, showing the influences of the parameters on the reflectance peaks, with the final parameter values and reflectance indicated by the white dashed lines in c-f.

### 2.2 Simulations of the Electric Near-Field Localizations and Enhancements

The plasmonic resonances arise from the localized enhanced electric near-field, which significantly strengthens the interaction between infrared light and analyte molecules. [34] Following the parameters optimization, we performed detailed near-field electromagnetic simulations using FDTD to analyze the field enhancement properties of ME-FIRST. As illustrated in **Figure 3**, the metasurface efficiently couples free-space incident waves into surface plasmon polaritons (SPPs), achieving strong field confinement across the fingerprint infrared region. When excited with linearly polarized infrared radiation at a 90° polarization angle, the strong coupling between the elementary radiative electric dipolar resonance of the center rod and the non-radiative fundamental magnetic modes of the side double 'N' leads to a mode hybridization, [35] producing two spectrally distinct but spatially overlapping plasmonic resonances at 1820 cm$^{-1}$ and 1300 cm$^{-1}$. In contrast, considering exciting the structure with linearly polarized infrared radiation at a 0° polarization angle, a strong SPP will only be generated at 1540 cm$^{-1}$, localized primarily within the double 'N' structure and filling the spectral gap between hybridized modes in 90° polarization. Numerical simulations reveal an average field enhancement factor $|E/E_0|$ of ~60-fold at the three reflectance peaks ($P_0$, $P_1$, $P_2$). Moreover, the plasmonic hot spots, concentrated at the rod ends and edges of the double 'N' structure, enable deep-subwavelength probing of biomolecular vibrations, making ME-FIRST highly sensitive to the same nanoscale analyte signatures.

To further quantify the field localization, we analyzed the spatial decay profile of the enhanced evanescent plasmonic fields localized around the hot spots of the Au resonator structures through numerical simulations. The enhancement factors exhibit an exponential decay with probing distance, maintaining significant light-matter interactions through electric near-field enhancement up to depths of ~100 nm above the metasurface for all three resonant modes ($P_0$, $P_1$, and $P_2$). This confined enhancement depth ensures efficient light-matter interaction across a biologically relevant sensing volume and reduces the absorption of water on IR light, which is critical for practical spectroscopic applications.

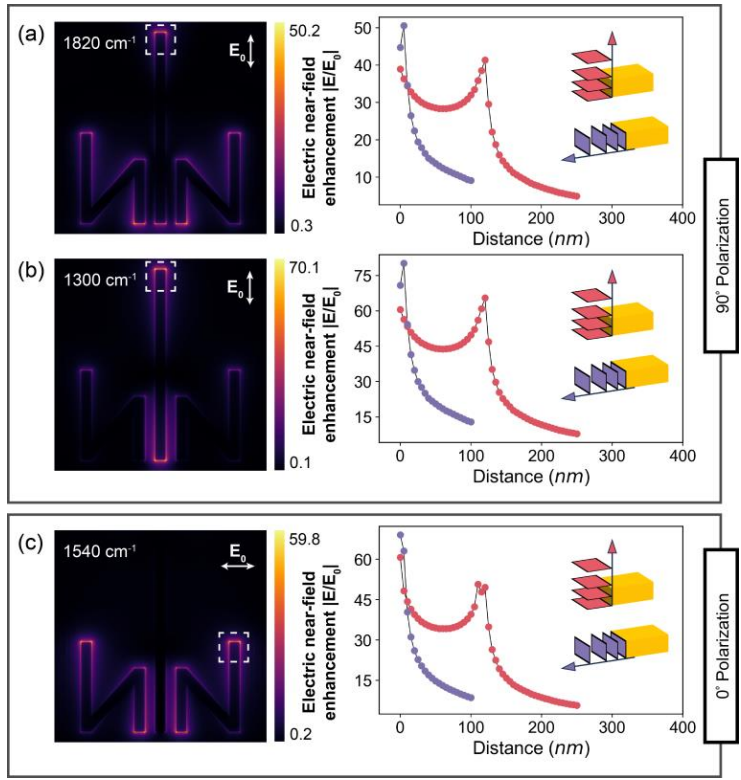

**Figure 3. Simulated electric near-field enhancement of ME-FIRST.** The electric near-field enhancement maps are shown on the left, and the plots of the spatial decay of the evanescent plasmonic resonance of the Au resonators enclosed with white dashed boxes at the corresponding resonance wavenumber are shown on the right. Simulated maximum E-field intensity enhancement extending out, perpendicular to the end face (purple markers, schematic) of the Au antenna or moving in the z-direction, through it (red markers, schematic). (a) The resonance at 1820 cm$^{-1}$. (b) The resonance at 1300 cm$^{-1}$. (a) and (b) are both excited with linearly polarized infrared light at a 90° polarization angle. (c) The resonance at 1540 cm$^{-1}$, excited with linearly polarized infrared light at a 0° polarization angle.

## 2.3 Fabrication and Validation of ME-FIRST

The ME-FIRST metasurface was fabricated on a CaF$_2$ substrate using a combination of electron-beam lithography (EBL) and electroplating techniques, with detailed fabrication protocols provided in Methods. Structural characterization via scanning electron microscopy (SEM) confirmed the successful realization of the designed metasurface, as evidenced by the images in **Figure 4**, which display the fabricated device at varying magnifications with geometric parameters matching those optimized in simulations (Figure 2a). **Figure 5a** shows the reflection infrared spectra of the fabricated ME-FIRST measured with incident light polarized at both 90° and 0° polarization angles, exhibiting excellent agreement with the simulated results from Figure 2b, with all three resonant modes observed at their predicted wavenumber positions. Moreover, the reflectance of ME-FIRST with the incident light at mixed polarization without any polarizer and analytes is shown in Figure 5a, approximately the mean of the spectra with 90° and 0° polarization.

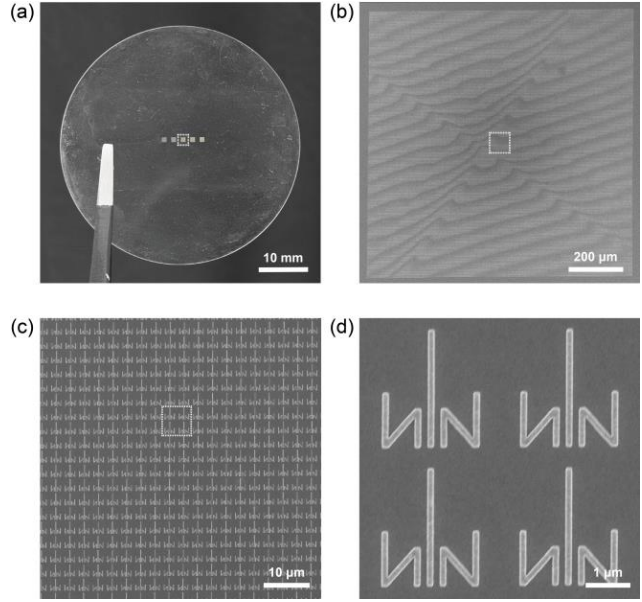

**Figure 4. Photograph and scanning electron microscopic (SEM) images of the fabricated ME-FIRST.** (a) Photograph image of the fabricated ME-FIRST. (b-d) SEM images of the fabricated ME-FIRST with different magnifications. The region enclosed by the white dashed box corresponds to the area of the image captured at a higher magnification.

To experimentally validate the performance of the ME-FIRST for the detection of fluid analytes, the water-immersed *L*-lysine amino acid is selected as the analyte for proof-of-concept. As a fundamental building block of proteins, amino acids are the most important macromolecules for the functions of humans and animals. *L*-lysine is one of the amino acids that are essential for human and animal nutrition, [36] making it a biologically relevant target for infrared spectroscopic analysis. We prepared a series of *L*-lysine solutions with six different concentrations of 5, 10, 20, 30, 40, and 50 milligrams per milliliter (mg/ml) to evaluate the performance of ME-FIRST in liquid environments. The absorbance spectrum $A(\omega)$ here is defined as [19]:

$$A(\omega) = -\log_{10}\frac{R_{L-lysine}}{R_{H_2O}} \quad (1)$$

, where $R_{L-lysine}$ is the measured reflectance spectra with the *L*-lysine solutions on the ME-FIRST, and $R_{H_2O}$ is the measured reflectance spectra with water on the ME-FIRST. In Figure 5b, we present the absorbance spectra of water-immersed *L*-lysine at different concentrations, as tested with ME-FIRST. The observed peaks correspond to characteristic vibrational modes of *L*-lysine shown in Table 1, [37,38], demonstrating ME-FIRST's ability to detect molecular fingerprints in aqueous environments.

**Table 1. Assignments for the absorption bands of dissolved *L*-lysine**

| Peak Position (cm$^{-1}$) | Assignment |
|---|---|
| 1598 cm$^{-1}$ | COO$^-$ asymmetric stretching |
|  | NH$^+_3$ asymmetric deformation |
| 1559 cm$^{-1}$ | NH$_2$ scissoring |
| 1521 cm$^{-1}$ | NH$^+_3$ symmetric deformation |
| 1412 cm$^{-1}$ | COO- symmetric stretching |
| 1351 cm$^{-1}$ | CH$_2$ out-of-plane wagging |
| 1151 cm$^{-1}$ | NH$^+_3$ rocking |

To quantify the detection capability of ME-FIRST, we performed integrated absorbance calibration across four spectral regions: 1650-1480 cm$^{-1}$, 1440-1380 cm$^{-1}$, 1380-1300 cm$^{-1}$, and 1200-1100 cm$^{-1}$, each encompassing several characteristic absorption peaks. From the calibration curves, the limit of detection (LOD) is calculated as: [39]

$$LOD = 3 \times (SD\ of\ blank) \times sensitivity^{-1} \quad (2)$$

where the SD of blank is the standard deviation of nine consecutive measurements on DI water, and sensitivity is the slope of the calibration curve. The most sensitive detection was achieved in the 1650-1480 cm$^{-1}$ range, where ME-FIRST demonstrates an LOD of 0.076 mg/ml for *L*-lysine with a sensitivity of 0.131 ml/mg, attributed to the strong vibrational modes in this spectral region. The error bars indicate the standard deviation of the mean values of nine measurements. The deviation is caused by experimental measurement error, including environmental conditions and random fluctuations of the light source. Despite these minor variations, the consistently high signal-to-noise ratio confirms that the employment of the designed metasurface is able to provide robust IR spectra.

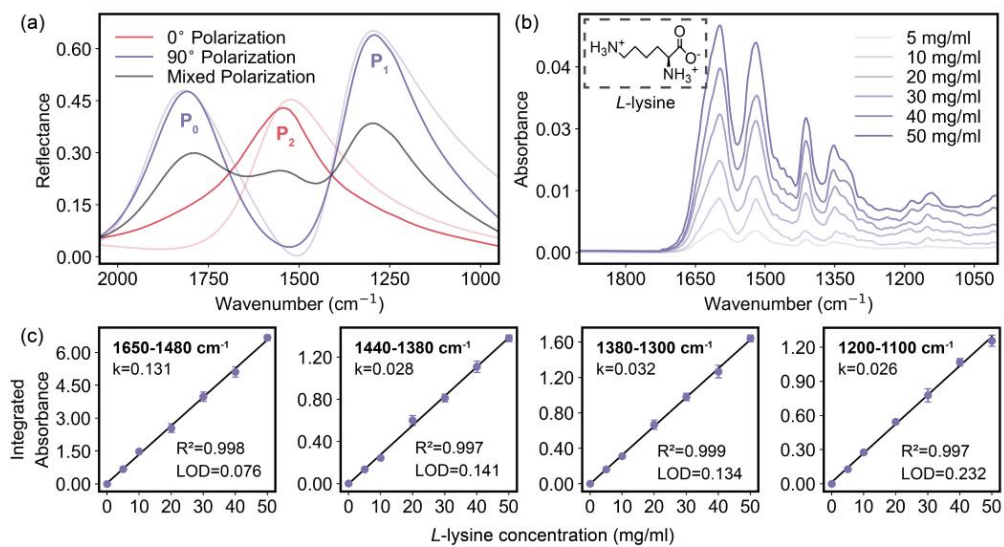

**Figure 5. Results of experimental measurement and validation of the fabricated ME-FIRST.** (a) The measured reflectance infrared spectra of the fabricated ME-FIRST. The spectra measured with incident light polarized at 90° and 0° polarization angles are shown with their corresponding simulated spectra represented with transparent shades of the same color. The reflectance spectrum of mixed polarization without a polarizer is depicted with the gray line. (b) Absorbance spectra for *L*-lysine in water with concentrations of 5, 10, 20, 30, 40, and 50 mg/ml. The molecular formula of an *L*-lysine molecule in water is shown. (c) The calibration curves

with the integrated absorbance of four spectral regions: 1650-1480 cm$^{-1}$, 1440-1380 cm$^{-1}$, 1380-1300 cm$^{-1}$, and 1200-1100 cm$^{-1}$, each encompassing several characteristic absorption peaks. The limit of detection (LOD) is expressed in mg/ml, and the sensitivity is expressed with the slope *k* in ml/mg.

Our proof-of-concept experimental results confirm that our ME-FIRST successfully acquires high-quality fingerprint-region spectra of liquid analytes with good sensitivity. To further validate the effective electric near-field enhancement effect of our ME-FIRST, we compared the results to a prior study testing *L*-lysine with a nanorod metasurface with single resonance promoted by *Mahalanabish, A*. et.al. [19] They designed a 3D nano-structure elevated by dielectric material with a sensitivity of 0.121 ml/mg, and a LOD of 0.10 mg/ml within our first segment of spectral integration, which has been demonstrated to surpass the performance of the same 2D sensor and ATR-FTIR in their previous work. [40] Considering the performance of ME-FIRST, we successfully detect the *L*-lysine with a sensitivity of 0.131 ml/mg and a LOD of 0.076 mg/ml in the same segment, reaching a comparable level of absorbance (a maximum of ~0.05 for *L*-lysine solution with a concentration of 50 mg/ml) and performance. Moreover, the simulated enhancement factors of ME-FIRST are at a similar level compared to those of the theoretical simulations from the metasurfaces in the earlier studies, which are used for *L*-lysine detection, [19] and metasurface-enhanced fingerprint infrared spectroscopy, [30] respectively, collectively confirming the effectiveness of our ME-FIRST platform.

Furthermore, several strategies can be implemented in future work to enhance the detection signal-to-noise ratio (SNR), reduce measurement errors, and ultimately lower the calculated LOD while maintaining full fingerprint spectral coverage. First, fabrication precision can be further improved to enhance agreement between experimental results and simulations, ensuring a more consistent performance of the device. Second, a higher-performance light source, such as a quantum cascade laser (QCL) or synchrotron light source, instead of the Globar light source in our FTIR spectrometer system, could be employed to provide stronger and more stable signal intensity, which will lead to a reduced deviation. Third, the height of the metasurface may be increased by incorporating dielectric materials with tailored thickness, [19] thereby expanding the interaction area between the metasurface and the analyte solution, thus enhancing the light-matter coupling.

## 3. Conclusion

In this study, we proposed a metasurface-enhanced fingerprint infrared spectroscopic tool (ME-FIRST) for highly sensitive infrared spectroscopy of fluid analytes. The metaunits design, featuring a central nanorod with a side double 'N' structure, generates three polarization-multiplexed plasmonic resonances. Thus, ME-FIRST can support spatially overlapping yet spectrally distinct plasmonic resonances across the full fingerprint infrared region (1900-1000 cm$^{-1}$), thus enhancing the light-matter interactions in sub-wavelength volumes and enabling the spectra measurement in liquid analytes. Numerical simulations reveal the formation of large-area hot spots with strong near-field confinement and enhancement, translating to superior sensing performance. Experimental validation demonstrates the performance of ME-FIRST for sensing the molecular vibrational modes in *L*-lysine, achieving a competitive sensitivity of 0.131 ml/mg and a limit of detection (LOD) of 0.076 mg/ml by the integrated absorbance in the 1650–1480 cm$^{-1}$ range while maintaining complete spectral coverage. The observed signal robustness, despite minor environmental and instrumental fluctuations, underscores the reliability of this metasurace-based fingerprint infrared detection platform. Beyond liquid analysis, ME-FIRST holds significant potential for solid-sample applications, including tissue histopathology, opening new avenues for disease diagnostics, prognostics, and therapeutic monitoring through label-free molecular fingerprinting.

## 4. Methods

*4.1 Electromagnetic Simulations*

Electromagnetic parameter-scanning simulations were conducted using the finite-difference time-domain (FDTD) solver from Lumerical. For numerical calculation of the reflectance and near-field enhancement, simulations were further performed by the FDTD solver. 3D simulations were carried out with periodic boundary conditions in the x-y plane and perfectly matched layer boundary conditions in the z direction, which was the direction of incident light propagation, with the maximum simulation time set as 5000 fs. The optical constants for the Au, Ti, and $CaF_2$ were taken from references. [42–44]

### 4.2 Nanofabrication of the Plasmonic Metasurface

Double-side-polished, single-crystal Calcium fluoride ($CaF_2$) rounds (50.8 mm diameter, 0.5 mm thickness) were used as substrates to fabricate the plasmonic metasurfaces of ME-FIRST. $CaF_2$ is an advantageous material for the fabrication of metasurfaces operating in the fingerprint region of the infrared spectrum. Its low absorption in this spectral range, combined with high optical transparency and a relatively low refractive index, allows for the efficient and accurate manipulation of infrared waves without significant losses. Additionally, $CaF_2$ exhibits excellent surface polish ability and mechanical stability, which are essential for achieving high-fidelity nanoscale patterning during the fabrication of subwavelength metasurfaces.

Electron Beam Lithography (EBL) is chosen for fabricating ME-FIRST in this work. EBL is a powerful nanofabrication technique that enables the precise patterning with sub-10 nm precision, offering exceptional resolution and high pattern fidelity, allowing for the accurate definition of nanostructures critical for controlling electromagnetic wave manipulation, making it particularly suitable for the fabrication of ME-FIRST.

Nanostructures were fabricated by EBL, electroplating, lift-off, and an ion beam etching process sequence. Sequentially, a thin layer of metallic seed (~ 5 nm Ti and 10 nm Au) was sputtered on the substrates. One 380 nm resist layer (ZEP 520A) was spin-coated at 6000 rpm and baked for 2 min at 180°C. Five metasurfaces of 1 × 1 mm$^2$ area were patterned with electron beams of 200, 230, 260, 290, 320 uc/cm$^2$, respectively, and the center pattern was chosen for the experiment. Following the resist exposure, the ZEP 520A was developed in N50 solution (for exposing ZEP 520A exclusively) for 60 sec and IPA for 30 sec. Au nanostructures were formed by electroplating a ≈100 nm-thick Au layer, followed by a lift-off process with ZDMAC (mainly dimethylacetamide). To remove any residues from the plasmonic metasurfaces, an extra oxygen plasma treatment step was necessary after the metal lift-off process. Finally, an ion beam etching process was carried out to remove the seed layer.

### 4.3 Infrared Spectroscopy Measurements

All the reflection spectra of the ME-FIRST sensor were measured under ambient conditions at room temperature, using a Fourier transform infrared spectrometer (Bruker VERTEX 80v), which was connected to an infrared microscope equipped with a KBr beam splitter and a liquid nitrogen-cooled MCT-D316 detector. The incident broadband infrared light from a Globar inside the FTIR was focused by a reflective objective of the microscope, and the reflected light was collected by the same objective. A wire grid polarizer was used to make the incident light polarized to test the reflectance on different polarization directions. The reference reflection spectra for calculating the reflectance were measured from a reference gold mirror. Each spectrum was measured with a spectral resolution of 4 cm$^{-1}$ and averaged over 32 scans, and measured on an area of 200 μm by 200 μm by setting the aperture of the infrared microscopy.

**Disclosure statement**

All the authors declared no competing interests.

**Author contributions**

Xiangyu Zhao: Conceptualization, Data Curation, Formal analysis, Investigation, Methodology, Validation, Visualization, Writing – original draft, Writing - Original Draft. Yuqing Liu:

Conceptualization, Methodology, Writing - Review & Editing. Jingzhu Shao: Writing - Review & Editing. Longsheng Fang: Validation, Writing - Review & Editing. Chongzhao Wu: Conceptualization, Methodology, Formal analysis, Supervision, Project administration, Funding acquisition, Writing - Review & Editing.

## Acknowledgments

The work is supported by the National Natural Science Foundation of China under Grant 62375170 and 62535019, the Shanghai Jiao Tong University under Grant YG2024QNA51, and the Science and Technology Commission of Shanghai Municipality under Grant 20DZ222040. We also thank the Shanghai Synchrotron Radiation Facility of BL06B (https://cstr.cn/31124.02.SSRF.BL06B) for the assistance.